\documentclass[aps,pra,a4paper,twocolumn,amsmath,amssymb,showpacs,superscriptaddress]{revtex4-1}

\usepackage{graphicx}% Include figure files
\usepackage{dcolumn}% Align table columns on decimal point
\usepackage{bm}% bold math

\begin{document}

%\preprint{Draft version}

\title{Van der Waals coefficients for systems with ultracold polar alkali-metal molecules}

\author{P. S. \.{Z}uchowski}
\email{\texttt{pzuch@fizyka.umk.pl}}
\affiliation{Institute of Physics, Faculty of Physics, Astronomy and Informatics, Nicolaus Copernicus University, Grudziadzka 5, 87-100 Torun, Poland}

\author{M. Kosicki}
\affiliation{Physics Institute, Kazimierz Wielki University, pl. Weysenhoffa 11, 85-072 Bydgoszcz, Poland}

\author{M. Kodrycka}
\affiliation{Physics Institute, Kazimierz Wielki University, pl. Weysenhoffa 11, 85-072 Bydgoszcz, Poland}

\author{P. Sold\'{a}n}
\email{corresponding author: \texttt{pavel.soldan@mff.cuni.cz}}
\affiliation{Department of Chemical Physics and Optics, Faculty of Mathematics and Physics, Charles University in Prague, Ke Karlovu 3, CZ-12116 Prague 2, Czech Republic}

\begin{abstract}
A systematic study of the leading isotropic van der Waals coefficients for the alkali-metal atom + molecule and molecule + molecule systems is presented. Dipole moments and static and dynamic dipole polarizabilities are calculated employing high-level quantum chemistry calculations. The dispersion, induction, and rotational parts of the isotropic van der Waals coefficient are evaluated. The known van der Waals coefficients are then used to derive characteristics essential for simple models of the collisions involving the corresponding ultracold polar molecules.
\end{abstract}



\maketitle
%\newpage

\section{Introduction}

Ultracold chemistry in sub-microkelvin regime has emerged as one of the most
exciting fields in atomic and molecular physics \cite{Hutson:IRPC:2006,Carr:2009,Ulmanis:2012}.
By tuning  magnetic field across Feshbach resonances
one can combine two free atoms into a bound state,
with binding energy of order of MHz and then, with elaborated laser techniques,
coherently transfer them into the deeply bound states - including the absolute
rotational-vibrational-electronic ground state. At present two  alkali-metal dimers have been produced
in this manner KRb and Cs$_2$ \cite{Ni:KRb:2008,Danzl:v73:2008}.
It is also worth mentioning that LiCs molecules in the vibrational ground state have been
produced by photoassociation followed by spontaneous emission \cite{Deiglmayr:2008}.
At present, many experimental groups have focused on production of other heteronuclear
alkali-metal dimers hoping to obtain ultracold quantum gases of polar molecules, stable with respect to
the atom exchange and trimer formation \cite{Zuchowski:2010a}. Such quantum gases of polar molecules will be
used to explore new ideas in quantum information theory \cite{DeMille:2002,RablPRL06}, quantum simulations of
condensed-phase physics \cite{BuechlerPRL07}, or fundamental studies of chemical reactions \cite{Ospelkaus:2009}.

Description of chemical processes in sub-microkelvin regime is extremely difficult,
because the full quantum calculation for such systems is nearly impossible.
Thus only few quantum dynamics studies of ultracold atom+diatom collisions employing global potential energy surfaces have been performed so far concentrating on the homonuclear spin-polarized systems \cite{Soldan:2002,Quemener:2004,Cvitas:bosefermi:2005,Cvitas:hetero:2005,Quemener:2005,Cvitas:li3:2007,Simoni:2009}, where single-electronic-state approach provides good approximation \cite{Hutson:IRPC:2007}.
Even then, the quantum dynamics calculations for heavier system are very challenging and have not been yet performed despite the fact that the corresponding quartet potential energy surfaces are rather simple \cite{Guerout:2009,Soldan:2010a}.

Theoretical treatment of non-spin-polarized systems would be even more challenging.
The calculations of  triatomic and tetraatomic  interaction potentials in such a case would have to include
many active electrons and coupled potential energy surfaces, which at present is very far from routine.
The following quantum dynamics calculations, especially in the presence of external fields, would be extremely demanding.
The interaction potentials involving alkali-metal atoms and dimers are likely to be strongly anisotropic, and therefore the basis sets for such
calculations would have to be very large.  On the other hand, there is a very small number of observables as outcome of ultracold collisions. After all, in a
laboratory we do not record state-resolved cross sections but only loss rates from
the state prepared before the experiment. Thus, the recent theories of ultracold collisions \cite{Idziaszek:2010:a,Idziaszek:2010,Kotochigova:2010,Quemener:2011},  formulated to explain
current experiments in this field, use only few simple  parameters that catch the essential physics.
Importantly enough, the  feature of the intermolecular interaction that
matters the most is the long-range shape of the interaction potential, usually represented analytically
by the well-known van der Waals expansion with the most important term  $-C_{6}R^{-6}$ ($R$ is the distance between the monomer centres of mass).

Properties of the alkali-metal dimers have been intensively studied using electronic structure methods.
A systematic study of the dipole moments of all possible alkali-metal dimers was published by
Aymar and Dulieu \cite{Aymar:2005}, and Deiglmeyr {\it et al.} \cite{Deiglmayr:2008pol} reported a systematic study of the static dipole polarizabilities for these systems.
Their approach, based on large effective core potentials combined with appropriately set core-polarization potentials, was particularly successful in
predicting binding energies and spectroscopic properties  of the alkali-metal dimers in the ground and low-lying excited states.

In this paper we report a systematic \textit{ab initio} study of the isotropic van der Waals $C_{6}$ coefficients for the alkali-metal atom + molecule (A+AB) and molecule + molecule (AB + AB) systems. We also derive characteristics essential for simple models of the corresponding ultra-low-energy collisions. In the following calculations masses of the bosonic $^{7}$Li, $^{23}$Na, $^{41}$K, $^{87}$Rb, and $^{133}$Cs isotopes were used.
% End of INTRODUCTION

\section{Methodology}

% --------------
% This part is being written now
%
The purpose of this paper is to provide essential parameters for
modelling of collisions between the polar molecules in their ground rovibrational state.
If colliding  molecules are in $j=0$ states only isotropic part of the interaction potential
governs its scattering properties at very long range -- larger than $R_{\rm vdW} =(2mC_6/\hbar^2 )^\frac{1}{4}$, where $m$ is the reduced mass of the colliding system.
If the strength of the  anisotropy of the potential becomes comparable with spacing of the  appropriate rotational energy levels of the molecule, then anisotropic term becomes
important: for example in case of A+AB collisions, the coupling driven by  $C_{62}$ between  $j=0$ and  $j=2$ channels becomes important if the potential anisotropy is comparable with $6B$. The same argument holds also for the AB+AB collisions.

It is well known \cite{Stone:2002,Kaplan:2006} that within the Born-Oppenheimer approximation the isotropic van der Waals $C_{6}$ coefficient of a two-monomer system (X+Y) can be decomposed into two contributions
each corresponding to a different effect.
First, it contains a \textit{dispersion} contribution $C_{6}^{\rm disp}$ that physically represents the interaction of fluctuating instantaneous
dipole moments, which are due to the movements of electrons, which correlate between interacting species at long range. Secondly,
in the case of the heteronuclear dimers, a permanent molecular dipole moment induces a dipole moment on the atom, which in turn interacts with the permanent molecular
dipole moment. This \textit{induction} contribution $C_{6}^{\rm ind}$ is usually smaller than the dispersion contribution.

The dispersion contribution to the isotropic van der Waals $C_{6}$ coefficient  can be calculated from the following integral
\begin{equation}
C^{\rm disp}_{6}  =\frac{3}{\pi}\int_0^{\infty} \bar{\alpha}_{\rm X}(i\omega) \bar{\alpha}_{\rm Y}(i\omega) d\omega
\label{disp}
\end{equation}
where $i$ is the unit imaginary number, $\omega$ is frequency, and
\begin{equation}
 \bar{\alpha}_{\rm mol}(i\omega)  =  \frac{1}{3}[\alpha_{\rm xx}(i\omega) +\alpha_{\rm yy}(i\omega) +\alpha_{\rm zz}(i\omega) ] \\
\label{averpol}
\end{equation}
is the orientation-averaged molecular dynamic dipole polarizability.
The induction contribution to the isotropic van der Waals $C_{6}$ coefficient can be expressed as
\begin{eqnarray}
C^{\rm ind}_{6} & = & \mu_X^{2} \bar{\alpha}_{\rm Y}(0) + \mu_Y^{2} \bar{\alpha}_{\rm X}(0),
\label{ind}
\end{eqnarray}
where $\mu$ is the corresponding permanent molecular dipole moment.
If the monomer X is an atom in the spherical-symmetry ground state and the overlap of the charge distribution of interacting species can be neglected, the first term in the above equation vanishes.
The total isotropic $C_6$ coefficient of the atom + molecule system is
a sum of the dispersion and induction contributions.

For molecule + molecule systems in their ground rotational state
there also exists a (non-Born-Oppenheimer) \textit{rotational} contribution to the effective
isotropic  $C_6$ resulting from a second-order coupling of the dipole-dipole
term \cite{Avdeenkov:2002,Barnett:2006,Buchachenko:2012}. It which has the form
$C_{6}^{\rm rot}= \mu^{4}/6B$
where $B$ is the molecule rotational constant.
Then the total isotropic $C_6$ coefficient for the molecule + molecule system is
a sum of the dispersion, induction, and rotational contributions.

Proper choice of the electron basis set is crucial for quantum chemistry calculations of the dipole moments and polarizabilities.
For lithium and sodium atoms we have used available core-valence correlation-consistent
basis sets cc-pCV5Z designed by Prascher \textit{et al.} \cite{Prascher:2011}, which we augmented by one set of diffuse functions.
Effective-core potentials (ECPs) with tailored valence basis sets  for heavy (K - Fr) alkali metal atoms have been optimized by Lim {\it et al.} \cite{Lim:2005}.
These ECPs are small-core type potentials, \textit{i.e.} the outermost 9 electrons are described explicitly.
To eliminate possible errors due to the basis incompleteness we have improved the original valence basis sets by adding  $g$ and $h$ functions.
and augmenting the basis sets by one set of diffuse functions. %  (additional basis functions for K, Rb, Cs are presented in the supplementary material).
These basis sets have been tested on the atomic static dipolar polarizabilities, which have been calculated with the spin-restricted open-shell coupled cluster method \cite{Cizek:1966}
with single, double and non-iterative triple excitations [RCCSD(T)] employing a finite-field approach.
In all cases the agreement with the reference values of Derevianko \textit{et al.} \cite{Derevianko:2010} was very good (the difference for Na was 2.2 a.u.; less than one atomic unit for other alkali metals).
%(see Table \ref{atomic_polarizabilities}).

All alkali-metal dimers in their ground electronic state $X^{1}\Sigma^{+}$
(near their equilibrium lengths) have their excited states
significantly separated in energy, thus we can properly describe them  by a single-reference
Slater determinant, which is ideal for using the  coupled cluster  approach \cite{Cizek:1966}.
For the molecular calculations we took the equilibrium distances, which were optimized by
the Paris group \cite{Aymar:2005,Deiglmayr:2008pol}.

For the molecular dynamic polarizability calculations we employed the time-independent coupled cluster
polarization propagator method in singles- and doubles approximation (TI-CCSD). This was introduced by Moszynski {\em et al.} \cite{Moszynski:2005} and
implemented in Molpro 2010.2 program \cite{MOLPRO_brief:2010}.
Several approximations to the full time-independent polarization propagator were discussed by Korona \textit{et al.} \cite{Korona:06}.
In our study, we used the so-called  CCSD(3) approximation of the TI-CCSD method, which is exact to the third order
of the electronic correlation operator. In benchmark calculations against the dynamic dipole  polarizabilities
based on the full-configuration-interaction response functions, the CCSD(3) approximation demonstrated systematically
a smaller error than the other approximations introduced there \cite{Moszynski:2005,Korona:06}.  Finally, in this paper we have used
a finite-field CCSD(T) approach in order to evaluate dipole moments and static dipole polarizabilities of the alkali-metal dimers. Such calculations were also needed
to verify the accuracy of the TI-CCSD dynamic polarizabilities.

\section{Results and discussion}

The dipole moments of the heteronuclear alkali-metal dimers calculated with the finite-filed CCSD(T) method as the first derivatives of energy with respect to
the electric field applied are collected in Table \ref{dipole_moments}).
The FF-CCSD(T) data are in a good agreement with those obtained by Aymar and Dulieu \cite{Aymar:2005},
with an error of at most 10\% for LiNa and  KRb molecules - note that for these species the dipole moment and
 charge separation between atoms is  significantly smaller than in other cases and the corresponding dipole moments are small.
The dipole moment we have obtained for KRb (0.62 D) is somewhat larger than the experimental value
(0.566$\pm$0.017 D \cite{Ni:KRb:2008}) and the value in Ref. \cite{Buchachenko:2012}.
For the sake of consistency we used our calculated values in further calculations.  In addition to the finite-field CCSD(T) values  we also calculated the finite-field CCSD values in order to check how important is the inclusion of triply excited configurations in calculations of the alkali-metal dimer dipole moments. The FF-CCSD values are also in good agreement with our reference FF-CCSD(T) data (the FF-CCSD values are systematically higher by 10\%).
 The fact that the triples contribution to the dipole moments is not too substantial indicates that the FF-CCSD(T) result might be very close to real
 values as the expansion of the molecular wavefunction in terms of number of excitations should converge rather quickly.

\begin{table*}
\caption{Dipole moments (in Debye), rotational constant (in cm$^{-1}$), the orientation-averaged molecular static dipole polarizability (in atomic units), and the anisotropy of the molecular static dipole polarizability (in atomic units) of the ground $X^{1}\Sigma^{+}$ states of heteronuclear alkali-metal dimers calculated  at the equilibrium interatomic distances from Ref. \cite{Aymar:2005}.}
\begin{ruledtabular}
\begin{tabular}{lllrrrrr}
dimer & $r_{e}/a_{0}$  \cite{Aymar:2005} & $B/hc$ & $\mu$ &  $\bar{\alpha}^{\rm FF}_{\rm mol}(0)$  & $\bar{\alpha}^{\rm TI}_{\rm mol}(0)$ & $\Delta{\alpha}^{\rm FF}_{\rm mol}(0)$ & $\Delta{\alpha}^{\rm TI}_{\rm mol}(0)$ \\
\colrule
LiNa &  5.4518 &  0.425 &  0.48   &     237.7  &  237.6  & 156.3 & 155.7 \\
LiK  &  6.268  &  0.293 &  3.41   &     324.2  &  326.9  & 234.5 & 240.7 \\
LiRb &  6.5    &  0.254 &  3.99   &     347.2  &  352.1  & 262.0 & 272.7 \\
LiCs &  6.93   &  0.218 &  5.39   &     391.9  &  399.1  & 317.8 & 333.1 \\
NaK  &  6.61   &  0.094 &  2.72   &     358.1  &  362.7  & 247.2 & 260.9 \\
NaRb &  6.88   &  0.070 &  3.31   &     387.1  &  393.9  & 279.2 & 299.7 \\
NaCs &  7.27   &  0.058 &  4.63   &     439.3  &  448.0  & 339.4 & 364.1 \\
KRb  &  7.688  &  0.037 &  0.62   &     523.5  &  532.3  & 367.6 & 409.5 \\
KCs  &  8.095  &  0.030 &  1.98   &     596.0  &  606.8  & 436.1 & 488.9 \\
RbCs &  8.366  &  0.017 &  1.32   &     638.6  &  653.0  & 462.1 & 531.1 \\
\end{tabular}
\end{ruledtabular}
\label{dipole_moments}
\end{table*}

In order to verify the quality of the molecular dynamic dipole polarizabilities calculated with TI-CCSD we performed
further tests by checking their values in the static limit against the polarizabilities calculated with the FF-CCSD(T) approach and literature data.
As the reference values we have used those published by Deiglmayr \textit{et al.} \cite{Deiglmayr:2008pol} who used a 2-electron full configuration interaction method with
carefully tailored large-core effective core potentials including core polarization potentials. This approach has proven to be accurate, for example,
in predicting experimental values of the dipole moments of KRb  \cite{Ni:KRb:2008}, LiCs \cite{Deiglmayr:LiCs:2010}, and transition dipole moments  RbCs \cite{Debatin:2011} .
We have also used for comparison the values of Urban and Sadlej \cite{Urban:1995}, which were obtained with entirely different approach -
using all-electron basis sets with scalar relativistic effects included.
Our finite-field results agree very well with the results from Refs. \cite{Urban:1995,Deiglmayr:2008pol};
our FF-CCSD(T) values of the orientation-averaged molecular static dipole polarizability are systematically right in between their values with differences not exceeding  6\%.
The agreement  between  our TI-CCSD and FF-CCSD(T) values of the orientation-averaged molecular static dipole polarizability is even better (see Table \ref{dipole_moments}).
 With the exception of LiNa, where the difference is indeed negligible, the TI-CCSD values are systematically higher than
 the FF-CCSD(T) values with the differences never exceeding 2.5\%. The anisotropy of the molecular static dipole polarizability $\Delta{\alpha}_{\rm mol}(0)$ exhibits the same tendency. With the exception of LiNa the TI-CCSD values are systematically higher than the FF-CCSD(T) values with the differences ranging from 2.5\% to 13\%.

In the evaluation of formulas (\ref{disp}) and (\ref{ind}) we used the TI-CCSD values of the molecular dynamic dipole polarizabilities, the FF-CCSD(T) values of the molecular dipole moments and molecular static dipole polarizabilities, and the values of the atomic static and dynamic polarizabilities from Ref. \cite{Derevianko:2010}.
The integral in Eq. (\ref{disp}) was evaluated using Gauss quadrature for 50 frequencies also provided by Derevianko \textit{et al.}
\cite{Derevianko:2010}.

\begin{table}
\caption{The isotropic $C_6$ van der Waals coefficients (in atomic units) for the alkali-metal A + AB systems.
The last column shows  the value based on pair-wise atom-atom additive model.}
\label{vdw_ad}
\begin{ruledtabular}
\begin{tabular}{llrrrr}
 atom  & dimer &  $C_{6}^{\rm disp}$ & $C_{6}^{\rm ind}$ & $C_{6}$  &     $C_{6}^{\rm add}$     \\
\colrule
Li & LiNa    &  2217                 &               6   &  2223  &     2856   \\
   & LiK     &  2885                 &             294   &  3179  &     3711   \\
   & LiRb    &  3098                 &             407   &  3505  &     3934   \\
   & LiCs    &  3452                 &             740   &  4192  &     4454   \\
   \hline
Na & LiNa    &  2358                 &               6   &  2364  &     3023   \\
   & NaK     &  3405                 &             187   &  3592  &     4003   \\
   & NaRb    &  3673                 &             275   &  3948  &     4239   \\
   & NaCs    &  4092                 &             539   &  4631  &     4783   \\
   \hline
K  & LiK     &  4821                 &             520   &  5341  &     6219   \\
   & NaK     &  5364                 &             334   &  5698  &     6344   \\
   & KRb     &  7428                 &              17   &  7445  &     8171   \\
   & KCs     &  8298                 &             175   &  8473  &     9056   \\
   \hline
Rb & LiRb    &  5688                 &             790   &  6478  &     7235   \\
   & NaRb    &  6357                 &             539   &  6896  &     7373   \\
   & KRb     &  8154                 &              19   &  8173  &     8964   \\
   & RbCs    &  9751                 &              87   &  9838  &    10353   \\
   \hline
Cs & LiCs    &  7652                 &            1803   &  9455  &     9911   \\
   & NaCs    &  8555                 &            1324   &  9879  &    10073   \\
   & KCs     & 10995                 &             242   & 11237  &    12005   \\
   & RbCs    & 11772                 &             110   & 11882  &    12509   \\
\end{tabular}
\end{ruledtabular}
\end{table}

\begin{table}
\caption{The isotropic $C_6$  van der Waals coefficients (in atomic units) for the alkali-metal AB+AB systems.}
\label{vdw_dd}
\begin{ruledtabular}
\begin{tabular}{llrrr}
 dimer &  $C_{6}^{\rm disp}$  &  $C_{6}^{\rm ind}$  &     $C_{6}^{\rm rot}$  &  $C_6$  \\
\colrule
 LiNa    &     3582  &    17  &      110 &      3709 \\
 LiK     &     6024  &  1167  &   404491 &    411682 \\
 LiRb    &     6963  &  1711  &   876031 &    884705 \\
 LiCs    &     8670  &  3520  &  3397216 &   3409406 \\
   \hline
 NaK     &     7461  &   820  &   508325 &    516606 \\
 NaRb    &     8696  &  1313  &  1497080 &   1507089 \\
 NaCs    &    10822  &  2916  &  6932958 &   6946696 \\
   \hline
 KRb     &    14202  &    62  &     3456 &     17720 \\
 KCs     &    17716  &   723  &   450681 &    469120 \\
   \hline
 RbCs    &    20301  &   345  &   160336 &    180982 \\
\end{tabular}
\end{ruledtabular}
\end{table}

Tables \ref{vdw_ad} and \ref{vdw_dd} contain the predicted
isotropic van der Waals $C_{6}$ coefficients for the A+AB and AB+AB systems, respectively.
In the case of the A+AB systems there is a very clear progression in increase of the $C_6$ coefficient
for both A and B from Li toward Cs. The induction contribution to $C_6$ is usually small;
only in the case of significantly polar LiCs, LiRb, NaCs, and NaRb molecules
it is within the range 10-23\%.
Our $C_6$ values for the K+KRb, Rb+KRb, Rb+RbCs, and Cs+RbCs systems are systematically larger than those reported by Kotochigova \cite{Kotochigova:2010} by
 8\%, 6\%, 35\%, and 41\%, respectively.
The result for KRb+atom is clearly in good agreement with the result of Kotochigova, however, the difference for
RbCs+atoms is significantly larger.  The dynamic polarizability in Ref. \cite{Kotochigova:2010}
has been obtained as a sum-over-state with appropriate transition dipole moments of the RbCs molecule. It is likely that this way the RbCs dynamic polarizability might have
been underestimated using such procedure by neglecting some contributions or underestimating the
continuum contribution.
Note also, that the induction contribution for atom+diatom has not been included in Ref. \cite{Kotochigova:2010}.

It was proposed recently  to approximate the $C_6$ coefficients by simply
adding the pairwise atom-atom $C_6$ coefficients \cite{Kotochigova:2010,Mayle:2012}.
Our calculations have verified this model as seemingly reasonably good for heavy
atoms (Cs and Rb) interacting with weakly polar molecules.
This nice agreement is, however, fortuitous, since this additive approximation includes only
dispersion and no induction . In this case, the approximation of the trimer dispersion
forces by simply adding them among dimers overcompensates the lack
of the induction interaction.

As expected, the effective isotropic $C_6$ coefficients for the AB+AB systems very strongly depend on the AB dipole moment.
Only the LiNa and KRb dimers with smallest dipole moments are dominated by electronic contribution to the $C_6$ coefficient,
in other cases rotational contribution completely dominates the total $C_6$ coefficient. For the AB+AB systems, there  is also a very distinct
pattern in increase of the electronic contribution similar to the A+AB systems. The KRb isotropic $C_6$ coefficient  is  higher by 10\% compared to
the value of Kotochigova \cite{Kotochigova:2010} and by 6\% with respect to the value given by Buchachenko and coworkers
\cite{Buchachenko:2012}.
Our results are in agreement with those reported by
Qu\'{e}m\'{e}ner {\em et al.} \cite{Quemener:2011} for the LiNa (difference of 4\%), LiK (20\%), LiRb (17\%), and LiCs (11\%) systems.
These values are very sensitive to the dipole moment and rotational constant of the molecule,
 thus even small differences in these characteristics can easily translate into a 20\% difference in the dominating rotational part of the $C_6$ coefficient.

Known $C_6$ coefficients allow us to determine the energy limits for single partial-wave scattering. The p-wave or d-wave
scattering starts to dominate if the collision energy is comparable to the appropriate centrifugal barrier heights:
for the  A+AB collisions it is the p-wave, while for the bosonic AB+AB collisions it is the d-wave scattering.
In Table \ref{characteristics} we have included the centrifugal barrier heights  for the A+AB scattering and
bosonic  AB+AB scattering. Their values approximately determine the single partial-wave regime.
The same table contains also the mean scattering lengths \cite{Gribakin:1993}, which
illustrate, in a sense, a characteristic length scale of the corresponding interaction potential.

\begin{table}
\caption{Heights (in $\mu K$) of p-wave  centrifugal barriers $V_{\rm p}$ for the A+AB systems  and  d-wave centrifugal barriers $V_{\rm d}$ for the AB+AB systems, respectively,
 with mean scattering lengths $\bar{a}$ (in $a_0$) for the corresponding collisions.}
\label{characteristics}
\begin{ruledtabular}
\begin{tabular}{llcc|lcc}
 atom  & dimer &    $V_{\rm p}$     & $\bar{a}$    &  dimer  &  $V_{\rm d}$  & $\bar{a}$  \\
\colrule
Li & LiNa &  2442  &       39  & LiNa  & 2293 &    30  \\
   & LiK  &  1844  &       44  & LiK   &  111 &   110  \\
   & LiRb &  1581  &       46  & LiRb  &   28 &   157  \\
   & LiCs &  1397  &       48  & LiCs  &    8 &   244  \\
\cline{1-7}
Na & LiNa &   684  &       49  & & & \\
   & NaK  &   380  &       58  & NaK   &   64 &   125  \\
   & NaRb &   300  &       61  & NaRb  &   16 &   187  \\
   & NaCs &   256  &       65  & NaCs  &    6 &   251  \\
\cline{1-7}
K  & LiK  &   221  &       68  & & & \\
   & NaK  &   177  &       71  & & & \\
   & KRb  &   112  &       81  & KRb   &  119 &    64  \\
   & KCs  &    95  &       85  & KCs   &   15 &   155  \\
\cline{1-7}
Rb & LiRb &    64  &       86  & & & \\
   & NaRb &    56  &       89  & & & \\
   & KRb  &    47  &       95  & & & \\
   & RbCs &    32  &      104  & RbCs  &   17 &   131  \\
\cline{1-7}
Cs & LiCs &    29  &      105  & & & \\
   & NaCs &    26  &      108  & & & \\
   & KCs  &    23  &      112  & & & \\
   & RbCs &    19  &      117  & & & \\
\end{tabular}
\end{ruledtabular}
\end{table}

\section{Conclusions}

In conclusion,  we have reported a complete \textit{ab initio} study of the isotropic $C_6$ van der Waals coefficients in
all possible  A+AB and AB+AB systems, where A and B are two distinct alkali-metal atoms and AB are molecules are in their ground state.
Given the rapid development of the field and many ongoing experiments with polar alkali-metal molecules, we expect that these results
will be beneficial for modelling their collisional properties, which are crucial for stability studies of the ultracold molecular dipolar gases in traps.	
In future studies we would like to pay increased attention to the role of anisotropy in ultracold collisions, and we would also like to exploit our results when constructing potential energy surfaces for various collisional systems.

\section{Acknowledgements}

P.S.Z., M.K., and M.K. acknowledge funding from  the Homing Plus programme (Project No. 2011-3/14) of the Foundation for Polish Science, which is co-financed by the European  Regional Development Fund of the European Union. We are also grateful for the computer time from the Wroclaw Centre for Networking and Supercomputing.

%\bibliography{my_refs}{}


\end{document}